\documentclass[aip,apl,twocolumn,superscriptaddress,reprint]{revtex4-1}
\usepackage{amssymb}
\usepackage{amsmath}
\usepackage{graphicx}
\usepackage{natbib}
\usepackage{epstopdf}
\usepackage{color}

\definecolor{new}{rgb}{.38,.6,.38}
\definecolor{old}{rgb}{1,0,0}

\begin{document}
\title{Split-Gate Cavity Coupler for Silicon Circuit Quantum Electrodynamics}

\author{F. Borjans}
\affiliation{Department of Physics, Princeton University, Princeton, New Jersey 08544, USA}
\author{X. Croot}
\affiliation{Department of Physics, Princeton University, Princeton, New Jersey 08544, USA}
\author{S. Putz}
\altaffiliation{Present address: Vienna Center for Quantum Science and Technology, Universit\"at Wien, Boltzmanngasse 5, 1090 Vienna, Austria}
\affiliation{Department of Physics, Princeton University, Princeton, New Jersey 08544, USA}
\author{X. Mi}
\altaffiliation{Present address: Google Inc., Santa Barbara, California 93117, USA}
\affiliation{Department of Physics, Princeton University, Princeton, New Jersey 08544, USA}
\author{S. M. Quinn}
\affiliation{HRL Laboratories LLC, 3011 Malibu Canyon Road, Malibu, California 90265, USA}
\author{A. Pan}
\affiliation{HRL Laboratories LLC, 3011 Malibu Canyon Road, Malibu, California 90265, USA}
\author{J. Kerckhoff}
\affiliation{HRL Laboratories LLC, 3011 Malibu Canyon Road, Malibu, California 90265, USA}
\author{E. J. Pritchett}
\altaffiliation{IBM Thomas J Watson Research Center, 1101 Kitchawan Road, Yorktown Heights, NY 10598, USA}
\affiliation{HRL Laboratories LLC, 3011 Malibu Canyon Road, Malibu, California 90265, USA}
\author{C. A. Jackson}
\affiliation{HRL Laboratories LLC, 3011 Malibu Canyon Road, Malibu, California 90265, USA}
\author{L. F. Edge}
\affiliation{HRL Laboratories LLC, 3011 Malibu Canyon Road, Malibu, California 90265, USA}
\author{R. S. Ross}
\affiliation{HRL Laboratories LLC, 3011 Malibu Canyon Road, Malibu, California 90265, USA}
\author{T. D.  Ladd}
\affiliation{HRL Laboratories LLC, 3011 Malibu Canyon Road, Malibu, California 90265, USA}
\author{M. G. Borselli}
\affiliation{HRL Laboratories LLC, 3011 Malibu Canyon Road, Malibu, California 90265, USA}
\author{M. F. Gyure}
\affiliation{HRL Laboratories LLC, 3011 Malibu Canyon Road, Malibu, California 90265, USA}
\author{J. R. Petta}
\affiliation{Department of Physics, Princeton University, Princeton, New Jersey 08544, USA}


\begin{abstract}
Coherent charge-photon and spin-photon coupling has recently been achieved in silicon double quantum dots (DQD). Here we demonstrate a versatile split-gate cavity-coupler that allows more than one DQD to be coupled to the same microwave cavity. Measurements of the cavity transmission as a function of level detuning yield a charge cavity coupling rate $g_\text{c} / 2\pi =$ 58 MHz, charge decoherence rate $\gamma_\text{c} / 2\pi =$ 36 MHz, and cavity decay rate $\kappa / 2\pi =$ 1.2 MHz. The charge cavity coupling rate is in good agreement with device simulations. Our coupling technique can be extended to enable simultaneous coupling of multiple DQDs to the same cavity mode, opening the door to long-range coupling of semiconductor qubits using microwave frequency photons.
\end{abstract}

\maketitle
Advances in Si/SiGe heterostructure growth \cite{Deelman2016} and the development of multilayer gate architectures that tightly confine quantum dot electrons \cite{angus2007,zajac2015reconfigurable} have led to a rapid rise of silicon-based quantum computing architectures. As predicted by theory \cite{Witzel2010}, isotopic enrichment of the semiconductor host material has been shown to greatly increase spin coherence times \cite{tyryshkin2012electron,steger2012}. Recently single qubit gates with fidelities exceeding 99.9\% have been demonstrated in isotopically enriched $^{28}$Si \cite{Yoneda2018}. Perhaps more importantly, silicon quantum devices now support competitive two qubit gate fidelities $>$98\% \cite{Huang2018,Xue2018}. However, two qubit gates using spins \cite{Veldhorst2015,Zajac2018,Watson2018} are generally based on the exchange interaction \cite{loss1998,petta2005} or capacitive coupling \cite{Shulman2012}. Both approaches are short-ranged, as exchange requires wavefunction overlap and coupling capacitances fall off rapidly with distance.

To achieve long-range qubit-qubit interactions, coupling of semiconductor qubits to the electromagnetic field of a superconducting cavity has been proposed \cite{childress2004,trif2008}. Coherent coupling of a single charge to a single photon has been demonstrated in Si \cite{Mi2017} and GaAs devices \cite{stock2017}. Moreover, coherent coupling of a single spin to a single photon has been achieved using electric-dipole coupling in combination with a synthetic spin-orbit field in Si \cite{Mi2018,Samk2018}. Alternative approaches using three electron spin states are being investigated in GaAs triple quantum dots \cite{landig2018}.

\begin{figure}[tbp]
	\centering
	\includegraphics[width=\columnwidth]{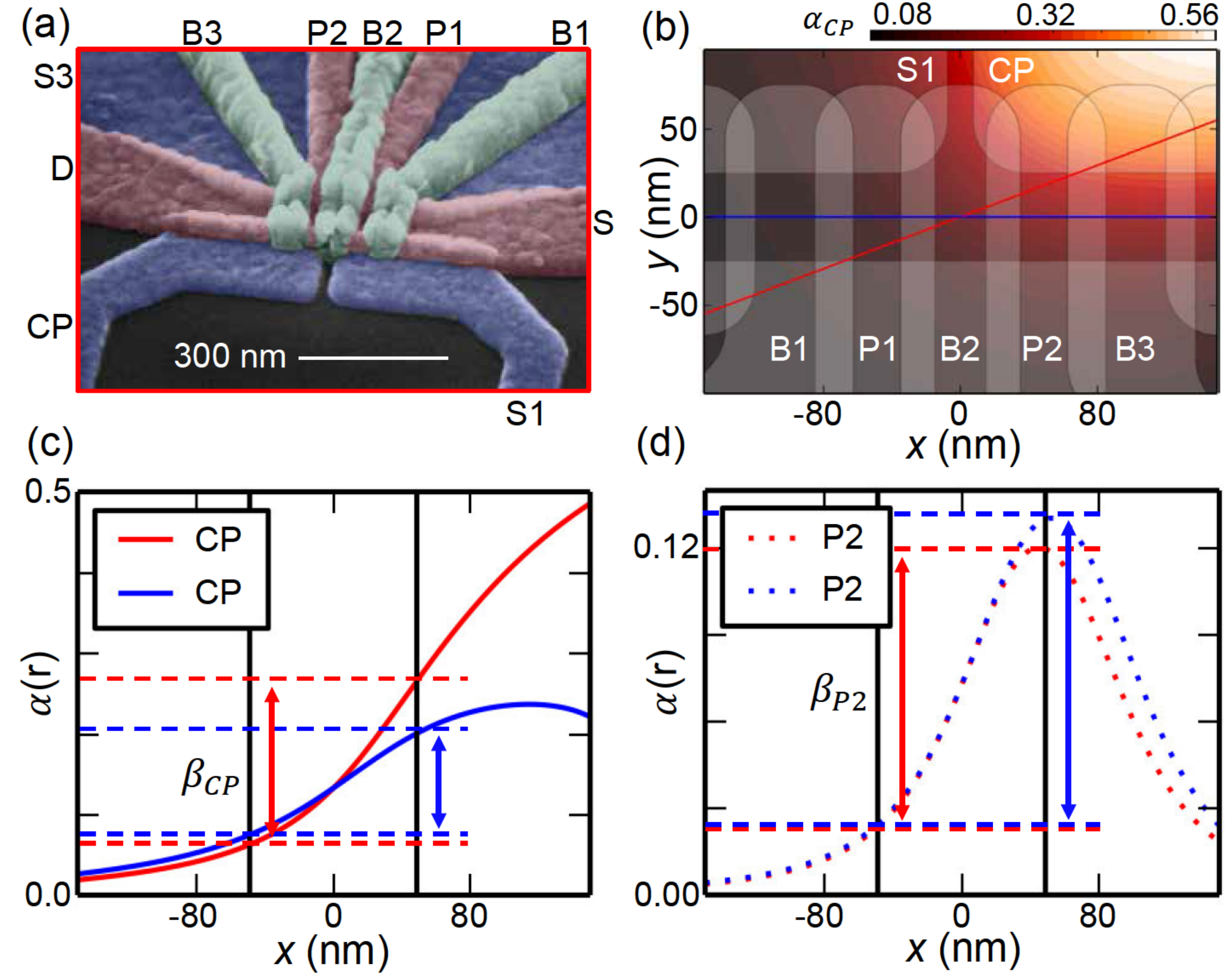}
	\caption{(a) False-color scanning electron microscope image of an accumulation-mode DQD. The DQD is formed beneath gates P1 and P2. The voltage on barrier gate B2, $V_{\rm B2}$, is used to tune the interdot tunnel coupling $t_{\rm c}$. (b) 2D spatial plot of the dimensionless lever arm $\alpha_{\rm CP}(r)=\tfrac{1}{e}\tfrac{\partial U_{well}}{\partial V_{\rm CP}}$ in the center of the Si quantum well with the projected gate stack overlaid. Two slices corresponding to possible DQD orientations are additionally indicated as red and blue lines. (c) Split-gate simulated lever arm $\alpha_{\rm CP}$ along the indicated red and blue slices. Vertical black lines indicate the approximate positions of electrons in the DQD; cavity coupling is proportional to $\beta_{\rm CP}$, the difference of $\alpha_{\rm CP}$ across the DQD. (d) Plunger gate lever arm $\alpha_{P2}(r)=\tfrac{1}{e}\tfrac{\partial U_{well}}{\partial V_{P2}}$ along the same slices, for comparison.}
	\label{fig:1}
\end{figure}

In many hybrid semiconductor-superconductor devices, coupling between the electric field of the superconducting cavity and the electron trapped in a DQD is achieved by galvanically coupling a plunger gate electrode of the DQD to the center pin of a superconducting coplanar waveguide cavity \cite{Delb2011,Frey2012,Toida2013,Deng2015}. This coupling approach has been very successful at the single qubit level, yielding charge $g_c/ 2\pi$ and spin $g_s/ 2\pi$ photon coupling rates large enough to reach the strong-coupling regime \cite{Mi2017,stock2017,Mi2018,Samk2018,landig2018}. However, since the electron occupancy of the DQD is adjusted using a plunger gate, which is in turn biased through the center-pin of the superconducting cavity, this approach has made it difficult to simultaneously couple two DQDs to the same cavity mode, as the two DQDs may each require very different plunger gate voltages for successful operation. In principle, device technology can be improved to obtain more consistent gate thresholds for single electron occupation \cite{Mills2018}. Another approach is to develop alternative cavity coupling techniques that do not rely on an actively tuned plunger gate. 

Here we demonstrate a split-gate cavity coupler for hybrid circuit quantum electrodynamics experiments that allows the coupling of more than one DQD to a common cavity mode. A $\lambda/2$ transmission line cavity with center frequency $f_c$ = 6.8 GHz is fabricated on a Si/SiGe heterostructure using electron beam lithography and reactive ion etching \cite{MiAPL2017}. An accumulation mode Si/SiGe DQD [Fig.\ 1(a)] is placed at each anti-node of the the cavity. The DQDs are fabricated using three overlapping layers of aluminum gates that are electrically isolated from one another by a native aluminum oxide barrier \cite{zajac2015reconfigurable}. These gates are protected from electrostatic discharge during packaging by ESD shorting bars, which are used to ground all of the gates. The shorting bars are opened after the sample has been wirebonded to the sample holder.

Coupling of an electron trapped in the DQD to the cavity mode is achieved using a split-gate architecture consisting of layer 1 screening gates S1 and CP [layer 1 is colored dark blue in Fig.\ 1(a)]. Gate CP is coupled to the center-pin of the superconducting cavity. With this device design it is feasible to couple more than one DQD to the same cavity mode since the screening gates allow substantially more bias tolerance in DQD tuning. Gates S and D are used to accumulate large source and drain Fermi reservoirs. DQD electrons are accumulated beneath plunger gates P1 and P2, while gate B1 (B3) is used to tune the tunnel coupling to the source (drain) reservoir. Finally, the voltage $V_{B2}$ on gate B2 is used to tune the interdot tunnel coupling $t_c$.

\begin{figure}[t]
	\centering
	\includegraphics[width=\columnwidth]{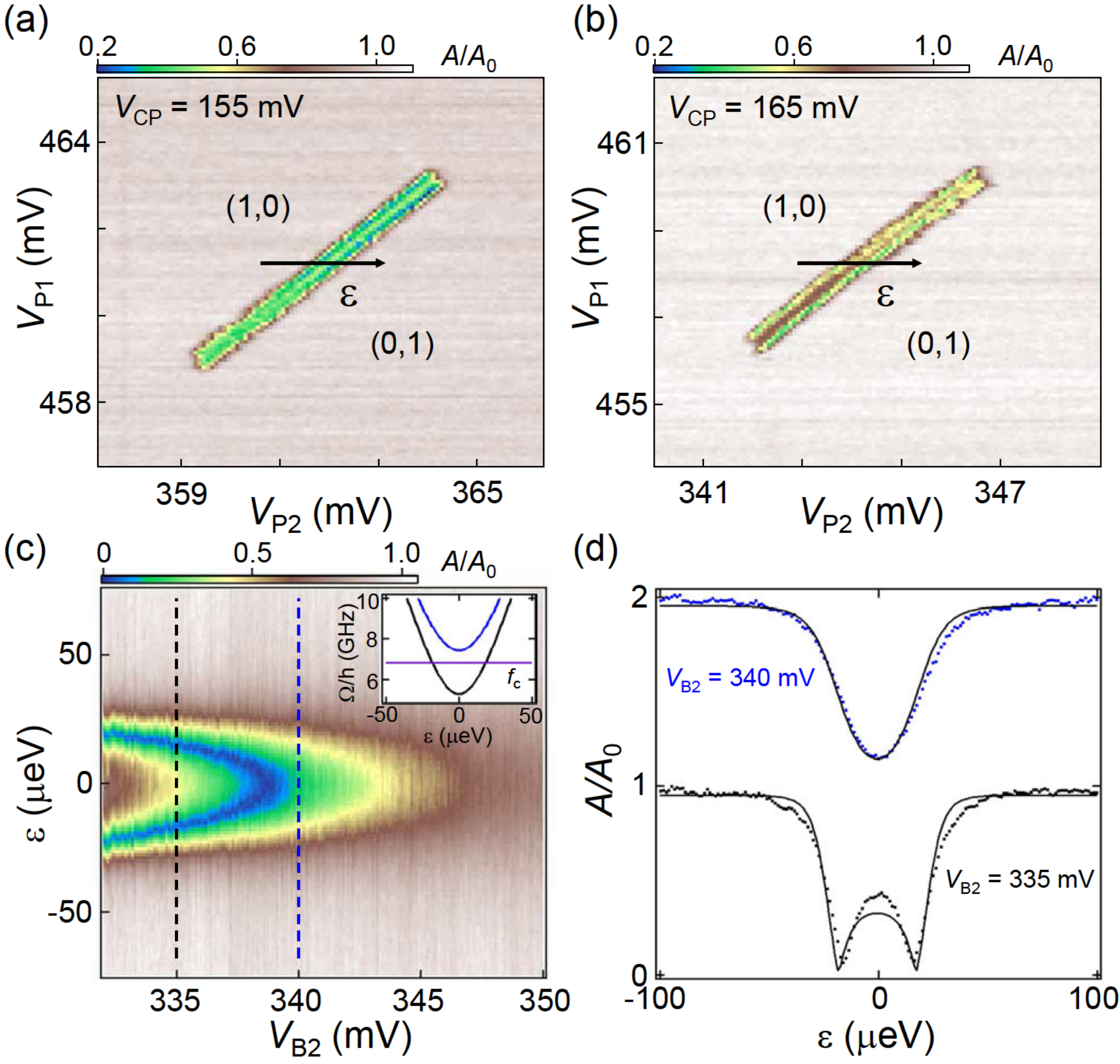}
	\caption{(a) $A/A_0$ measured near the (1,0)--(0,1) interdot charge transition with $V_{\rm CP}=155\,\rm mV$. (b) The interdot transition is tuned independently from the cavity gate CP, now with $V_{\rm CP}=165\,\rm mV$, using $V_{\rm P1}$ and $V_{\rm P2}$ [note the different axis ranges]. (c) $A/A_0$ plotted as a function of $\epsilon$ and $V_{\rm B2}$.  Inset: DQD transition frequency $\Omega/h$ plotted as a function of $\epsilon$ for $V_{\rm B2}$ = 335 mV and $V_{\rm B2}$ = 340 mV. The cavity resonance frequency $f_c$ is shown for comparison. (d) $A/A_0$ as a function of $\epsilon$ for $V_{\rm B2}$ = 335 mV and $V_{\rm B2}$ = 340 mV. These datasets are offset by 1 for clarity. Fits to the data (solid lines) yield $2t_c/h$ = 5.275 GHz and $2t_c/h$ = 7.432 GHz, respectively.}
	\label{fig:2}
\end{figure}

The benefits of the split-gate design can be further understood from electrostatic modeling. The cavity coupling rate of a DQD at zero detuning ($\epsilon$ = 0) is theoretically given by $g_c/2\pi=\frac{\beta}{2}f_c\sqrt{\frac{Z_0}{\pi\hbar}}$, where $Z_0$ is the cavity impedance and $\beta=e(\alpha_1-\alpha_2)$ is the differential lever arm of the microwave coupled gate across the DQD. Figure 1(b) depicts the dimensionless lever arm function $\alpha_{\rm CP}(r)=\frac{1}{e}\frac{\partial U_{well}}{\partial V_{\rm CP}}$ which is the spatial change in the potential of the quantum well due to the CP gate; this is obtained from self-consistent 3-D electrostatic calculations including the full gate layout as well as the accumulated 2DEG in the source/drain regions. The asymmetry of the split-gate design leads to large fringing fields across P1 and P2, hence the drastic variation in $\alpha_{\rm CP}$ within the active region. The dot-cavity coupling is maximized by aligning the DQD along these gradients; in a real device, the exact dot placement will be set by details of the tune-up and/or disorder. Figure 1(c) shows how the lever arm varies along two possible DQD orientations; $\beta_{\rm CP}$ increases from roughly 0.13 to 0.2 as the alignment improves. Assuming $Z_0=133\,\rm\Omega$ and $f_c=6.8\,\rm GHz$, this corresponds to $g_c/2\pi$ between $45-70\,\rm MHz$. In contrast, Fig.\ 1(d) shows the corresponding lever arm functions $\alpha_{P2}$ for the plunger gate P2, which are both smaller and less orientation sensitive due to that gate’s proximity to the DQD; the largest $\beta_{P2}= 0.11$, corresponding to $g_c/2\pi=38\,\rm MHz$. Thus, in addition to allowing multiple DQDs to address the cavity, the split-gate can lead to stronger cavity coupling, because of its large capacitance and fringing fields.

\begin{figure*}[t]
	\centering
	\includegraphics[width=1.75\columnwidth]{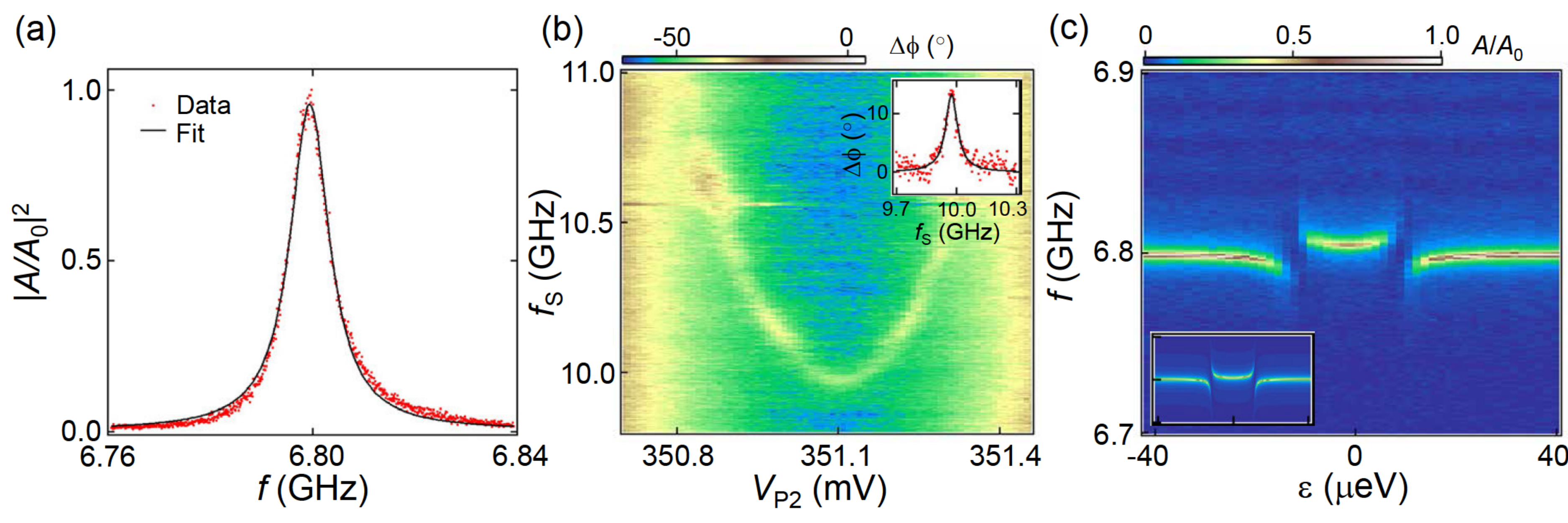}
	\caption{Device figures of merit. (a) Cavity transmission $|A/A_0|^2$ measured as a function of $f$. A fit to the data (solid line) yields $\kappa/2\pi$ = 1.2 MHz. (b) Cavity phase shift $\Delta \phi$ as a function of $\epsilon$ and spectroscopy tone frequency $f_s$ reveal the qubit dispersion relation $\Omega(\epsilon)/h$. Inset: The qubit transition linewidth yields a charge dephasing rate $\gamma_c/2\pi$ = 36 MHz. (c) Cavity transmission amplitude $A/A_0$ as a function of $f$ and $\epsilon$. Avoided crossings occur when $\Omega(\epsilon)/h$ crosses through the cavity frequency $f_c$. Inset: Theory prediction for $A/A_0$ with $\kappa/2\pi$ = 1.2 MHz,  $\gamma_c/2\pi$ = 36 MHz, and $g_c/2\pi$ = 50 MHz.}
	\label{fig:3}
\end{figure*}

The (1,0)-(0,1) interdot transition of the DQD is mapped out by measuring the cavity transmission amplitude $A/A_0$ as a function of plunger gate voltages $V_{P1}$ and $V_{P2}$ in Fig.\ 2(a). Here ($N_1$,$N_2$) refers to the number of electrons in dots 1 and 2. A strong reduction in cavity transmission $A/A_0$ is observed around $\epsilon$ = 0, which indicates substantial electric-dipole coupling of the DQD electron to cavity photons \cite{Delb2011,Frey2012,petersson2012}. Figure\ 2(b) shows the same DQD transition with the cavity voltage $V_{\rm CP}$ increased by $10\,\rm mV$. To maintain the same electron population, the two plunger gate voltages $V_{P1}$ and $V_{P2}$ are independently adjusted. As expected, $V_{P2}$ is more strongly affected than $V_{P1}$, confirming the desired difference in lever arms between the CP gate and the two dots.

To more quantitatively estimate the charge-cavity coupling rate $g_c$, we measure $A/A_0$ as a function of $\epsilon$ and $V_{\rm B2}$ in Fig.\ 2(c). For $V_{\rm B2}$ $>$ 345 mV, the charge qubit transition frequency $\Omega/h$ = $\sqrt{\epsilon^2 + 4 t_c^2}/h$ exceeds the cavity frequency $f_c$ for all values of $\epsilon$ ($h$ is Planck's constant). Here the dispersive shift is small and $A/A_0$ $\sim$ 1. As $V_{\rm B2}$ is reduced, e.g.\ around $V_{\rm B2}$ = 340 mV, the minimum charge qubit transition frequency $2 t_c/h$ approaches $f_c$ and there is a substantial reduction in $A/A_0$ near $\epsilon$ = 0. A further reduction in $V_{\rm B2}$ will pull $2 t_c/h$ below $f_c$. In this regime, e.g.\ with $V_{\rm B2}$ = 335 mV, the charge transition frequency will equal $f_c$ at two values of detuning that are symmetrically located about $\epsilon$ = 0. A large amplitude response will be observed near these values of detuning. The data in Fig.\ 2(d) show linecuts extracted from Fig.\ 2(c) at $V_{\rm B2}$ = 335 mV and $V_{\rm B2}$ = 340 mV. Fits to these curves yield $g_c/2\pi$ = 58 MHz and $2t_c/h$ = 5.275 GHz (7.432 GHz) for $V_{\rm B2}$ = 335 mV (340 mV).

We next compare the charge-photon coupling rate $g_c$ to the cavity loss rate $\kappa$ and charge dephasing rate $\gamma_c$ to assess the prospect of achieving strong charge-photon coupling with the split-gate cavity coupler. In Fig.\ 3(a) we plot the cavity transmission $|A/A_0|^2$ as a function of frequency $f$. Resonance is observed at $f$ = $f_c$ = 6.8 GHz and a cavity loss rate $\kappa/2\pi$ = 1.2 MHz is extracted from a Lorentzian fit to the data. The coherence of the charge qubit is probed using microwave spectroscopy. Here the cavity phase shift $\Delta \phi$ is measured using a weak probe tone at $f$ = $f_c$, while a spectroscopy tone of varying frequency $f_s$ is applied to the device. These data are shown in Fig.\ 3(b) and directly map out the qubit dispersion relation $\Omega(\epsilon)/h$. Fitting a linecut through these data at $\epsilon$ = 0 yields a charge dephasing rate $\gamma_c/2\pi$ = 36 MHz.

Lastly, we search for vacuum Rabi splitting in the cavity transmission amplitude as the charge qubit is brought into resonance with the cavity mode. Figure 3(c) shows the cavity transmission amplitude $A/A_0$ as a function of frequency $f$ and detuning $\epsilon$ with $2t_c$ = 6.2 GHz $<$ $f_c$. For this value of $t_c$, $\sqrt{\epsilon^2 + 4 t_c^2}$ = $hf_c$ when $\epsilon$ = $\pm$11.7~$\mu$eV. At these values of detuning, coherent coupling of the charge trapped in the DQD with the cavity mode leads to the vacuum Rabi splitting observed in the data. For comparison, the inset of Fig.\ 3(c) shows simulations of $A/A_0$ with $2t_c$ = 6.2 GHz, $g_c/2\pi$ = 50 MHz, and $\kappa/2\pi$ = 1.2 MHz.

In conclusion, we have developed a split-gate cavity coupler suitable for circuit quantum electrodynamics experiments with accumulation-mode Si DQDs. The accumulation gate design yields a well-defined DQD confinement potential that is capable of reaching the single electron regime. A substantial charge-cavity coupling rate $g_c/2\pi$ = 58 MHz is achieved, exceeding both the charge dephasing rate $\gamma_c/2\pi$ = 36 MHz and cavity loss rate $\kappa/2\pi$ = 1.2 MHz. Since the cavity coupler gate is not actively biased, our coupling approach can enable coupling of multiple DQDs to a common cavity mode, as recently demonstrated with two Si/SiGe spin qubits \cite{borjans2019}.

\begin{acknowledgements}
Supported by Army Research Office grant W911NF-15-1-0149 and the Gordon and Betty Moore Foundation's EPiQS Initiative through grant GBMF4535. Devices were fabricated in the Princeton University Quantum Device Nanofabrication Laboratory. The authors acknowledge the use of Princeton’s Imaging and Analysis Center, which is partially supported by the Princeton Center for Complex Materials, a National Science Foundation MRSEC program (DMR-1420541).
\end{acknowledgements}

\bibliographystyle{apsrev4-1}
\bibliography{Petta_APL_v2}

\end{document}